\documentclass[runningheads]{llncs}
\usepackage[T1]{fontenc}
\usepackage{graphicx}
\usepackage{xcolor}
\usepackage{hyperref}
\usepackage{amsmath}
\usepackage{orcidlink}
\begin{document}
\title{Are LLM-based Chatbots Good Enough to Support Computer Science Students in Multiple-Choice Exercises?}
\titlerunning{Are LLM-based Chatbots Good Enough for MCQ Support?}
%
\author{Markos Stamatakis\inst{1}\orcidlink{0000-0002-7974-308X} \and
Omkar Gavali\inst{2,3}\orcidlink{0009-0003-4735-6469} \and
Joshua Berger\inst{4}\orcidlink{0009-0001-0750-4143} \and
Christian Wartena\inst{4}\orcidlink{0000-0001-5483-1529} \and
Anett Hoppe\inst{1,2,3}\orcidlink{0000-0002-1452-9509} \and
Ralph Ewerth\inst{1,2,3,5}\orcidlink{0000-0003-0918-6297}
}
\authorrunning{M. Stamatakis et al.}
%
\institute{TIB -- Leibniz Information Centre for Science and Technology, Hannover, Germany
\and
University of Marburg, Marburg, Germany
\and
Hessian Center for AI (hessian.AI), Darmstadt/Marburg, Germany
\and
Hochschule Hannover -- Data$|$H Institute for Applied Data Science, Hannover, Germany 
\and
L3S Research Center -- Leibniz University Hannover, Hannover, Germany\\
\email{markos.stamatakis@tib.eu}}
%
\maketitle              
\begin{abstract}
Chatbots based on large language models (LLMs) are increasingly adopted for information retrieval, text generation, and writing assistance.
In educational settings, their use is also rapidly increasing.
Students leverage these systems to complete tasks, access information, and support learning.
However, the role of LLM-based chatbots in supporting learning and assessment in university-level computer science education is still underexplored.
To address this gap, we investigate the performance of several LLM-based chatbots in solving multiple-choice questions (MCQs) at the university level and evaluate their capabilities to assist student learning.
We developed 70 MCQs for a university lecture on interactive visual data analysis 
and evaluated the chatbots' performance using different prompt designs.
We further compared the results with students' performance. 
Finally, we conducted a user study in two
lectures (interactive visual data analysis, computer vision) to investigate how chatbot-generated answers and explanations affect students' performance.
The chatbot performance showed significant differences between smaller models and GPT-4o and GPT-5 models, which achieved the best results.
The results of the user study show that presenting ChatGPT answers together with an explanation does not improve students' performance in general. 
\end{abstract}
\section{Introduction}
Using LLM-based chatbots such as ChatGPT~\cite{openai2023gpt} and DeepSeek~\cite{deepseekr1} is becoming increasingly common in everyday tasks.
These models have proven effective for a wide range of applications, including text generation, summarization, translation, and problem-solving~\cite{llm_feature_survey}.
In education specifically, they are used increasingly to support both studying and teaching~\cite{chatgpt_chances_education,llm_chances_education}. 
Initial research suggests potential learning benefits:
students using an LLM-based system demonstrated higher knowledge gains than those using traditional textbooks~\cite{potential_learning_outcomes}. 
Studies have also explored LLMs for generating learning materials and assessment items at varying competency levels~\cite{school_level_questions,bloom_qg}, but less is known about their accuracy in answering domain-specific questions, despite their growing use as study aids.

In this paper, we address this gap and investigate whether LLM-based chatbots can accurately answer multiple-choice questions (MCQs) in a university-level computer science (CS) lecture on interactive visual data analysis (IVDA).
We manually created $70$ MCQs based on course slides as our evaluation dataset.
Specifically, we examine chatbot performance across different prompt variations using a structured prompting approach.
Additionally, we report results of a user study conducted in the lectures on IVDA and computer vision (CV) that investigated whether GPT-generated answer suggestions influence student behavior during assessments.
Drawing on the study results, we position the use of such systems within pedagogical and socio-technical perspectives.
We focus on the following research questions:
\begin{itemize}
    \item[\textbf{RQ1}] To what extent can LLM-based chatbots generate accurate answers to MCQs derived from a university-level IVDA lecture?
    \item[\textbf{RQ2}] How does prompt design affect the accuracy and reliability of LLM outputs on these MCQs?
    \item[\textbf{RQ3}] How do LLM-based outputs compare to student responses on these MCQs?
     \item[\textbf{RQ4}] How do GPT-generated answer suggestions influence student response patterns on MCQs in IVDA and CV?
\end{itemize}
This paper is structured as follows:
Sec.~\ref{sec:rw} reviews related work on question generation, answering, and the use of LLM-based chatbots in education.
Sec.~\ref{sec:method} describes the experimental setup.
Sec.~\ref{exp:answer_analysis} and \ref{exp:user_study} describe the evaluation of LLM performance on MCQs and the impact of GPT-generated outputs on student responses, respectively. 
Finally, Sec.~\ref{sec:conclusions} concludes and outlines future work.
\section{Related Work}
\label{sec:rw}
This section examines current research on automatic LLM-based question and answer generation (Sec.~\ref{sec:rw_qga}), and their use as a learning resource in educational contexts 
(Sec.~\ref{sec:rw_learning_resource}) to highlight challenges in analyzing models’ processing.
It is important to distinguish between direct LLM outputs and LLM-based chatbots. 
While LLMs generate outputs directly from prompts, chatbots add an interaction layer with predefined personas, instructions, or conversation management that shape processing.
\subsection{LLM-Based Question Generation and Answering}
\label{sec:rw_qga}
Prior work shows that models such as GPT-3.5 and GPT-4 can generate domain-specific questions and answers in zero-shot settings across fields like medicine, biology, and mathematics~\cite{gpt_microbiology,gpt_math,quiz_generation_gpt}, highlighting their potential for assessment and learning support. 
However, limitations in 
quality and reliability of outputs are still reported.
Generated questions and answers often require human verification due to duplicated output or misaligned difficulty level~\cite{gpt_problems_exam,quiz_generation_gpt}.
Furthermore, LLM hallucinations are a well-known problem~\cite{bang-etal-2025-hallulens,kalai2025languagemodelshallucinate}.
Such models can also be very sensitive to 
prompt formatting~\cite{promptformattingimpact},
and phrasing~\cite{robustnessreliability}. 
Focusing more on the nature of errors produced by LLM-based models in education, Liu et al.~\cite{llm_mistakes} analyzed the similarity of misconceptions between models and students' answers across subjects, finding moderate correlations that suggest partial alignment. 
Smaller models show a higher similarity to student mistakes, while larger models do not necessarily align with student error patterns.

These studies provide insights into the performance and limitations of LLM-based models in question generation and answering, but overlook the impact of prompting on reliability, such as the influence of persona, additional context, and response format.
Different domains were analyzed, but higher-level CS education is not the main focus.
We address this gap by evaluating $70$ manually designed MCQs across different models (Sec.~\ref{exp:answer_analysis}).

\subsection{LLM-based Chatbots for Learning} 
\label{sec:rw_learning_resource}
Common educational applications of LLM-based models include automated question generation, answer grading, code correction, and content explanation~\cite{llm_tutor_review}. 
Integrating these into learning environments presents both technical and pedagogical challenges.
Technical limitations include solving complex math problems and helping in programming tasks~\cite {review_llm_education}, model costs, and adaptability~\cite{shahzad2025comprehensive}.
Pedagogically, educators face challenges such as distinguishing LLM-generated work from student output and guiding appropriate use~\cite{shahzad2025comprehensive}.
Beyond the technical limitations, recent studies have begun to examine the impact of utilizing LLMs for learning.
For instance, research in programming education suggests that extensive reliance on LLM-based assistance may negatively affect final grades and decrease independent problem-solving abilities of undergraduate students~\cite{jovst2024impact}.
In contrast, research shows that LLM-based models deepen understanding of course content when used as teaching assistants rather than for answer retrieval~\cite{student_interaction_llm}.

Overall, the current research reflects that LLM-based models offer promising support for learning activities, but still have limitations in terms of deep understanding and problem-solving.
In particular, these studies do not reveal the extent to which the information provided by LLMs influences learners in test scenarios.
Therefore, we design a study to investigate how the availability of LLM-generated output on tests impacts learners in different scenarios (Sec.~\ref{exp:user_study}).
\section{Experimental Setup}
\label{sec:method}
This section describes our experimental approach.
The code and experimental data are available in a GitHub repository\footnote{\url{https://github.com/markossta/mcq_llm_answer_paper}}.
Sec.~\ref{sec:method_data} describes our MCQ dataset. We then introduce the investigated models (Sec.~\ref{sec:mcq_answering_models}), our prompting strategy (Sec.~\ref{sec:mcq_answering_prompting}), the user study design (Sec.~\ref{method:user_study}), and our evaluation methods (Sec.~\ref{method:evaluation}). 
\subsection{Multiple-Choice Questions (MCQs)}
\label{sec:method_data}
For evaluating LLM-based models' performance on generating answers for MCQs, we created $70$ MCQs based on a university-level CS lecture on IVDA. 
The lecture covers definitions, criteria, 
and visualization and analysis approaches for different data types, delivered through slide presentations~\cite{ivda_slides} under a CC-BY-4.0 license. 
We manually composed up to nine questions per presentation to test content understanding.
We selected IVDA as a less common CS topic, likely underrepresented in training data, to evaluate how well LLMs handle specialized content when answering MCQs in CS.
To ensure quality, we applied a two-step validation process: one author created the questions, which were then reviewed and discussed with a second author. Each question includes four answer options with a varying number of correct answers to prevent consistent response patterns. 
Correct answers are distributed evenly across all answer positions, 
preventing participants from benefiting from systematic guessing.
These questions were answered by up to $36$ students throughout multiple lectures, with a fixed set of MCQs assigned to each lecture unit.
Each set of MCQs referred to the topic of the previous week, encouraging students to revisit the material and enabling an assessment of their longer-term retention of the learned content.
All the answers of each student were collected anonymously with students’ consent, using only a participant ID for tracking the points.

For the user study, we used two sets of questions across two independent lectures to allow a comparison across different CS topics.
In the first lecture, the question set contains five of our created questions in the context of IVDA.
The second lecture focused on computer vision (CV), for which we created eight additional CV-MCQs, extending the already created $70$ IVDA-MCQs.
These questions are based on Chapter~9 of \emph{Foundations of Computer Vision}~\cite{foundations_cv_ch9} (CC-BY-ND-NC license) and include content related to how the learning process is performed to train models for solving specific problems.
We selected CV as a second scenario to examine how GPT-generated output of a more common topic affects student behavior and performance differently.
Both question sets include questions that GPT did not answer correctly, allowing analysis of how student answer patterns are impacted by correct and incorrect chatbot-generated outputs.
\subsection{Investigated Models}
\label{sec:mcq_answering_models}
To enable a comparison between smaller and bigger models, we included DeepSeek-R1~\cite{deepseekr1} and Mistral-7B~\cite{mistral_7b} in addition to ChatGPT~\cite{openai2023gpt} for evaluation.
For ChatGPT, we used GPT-4o and GPT-5.2 via the API connection.
We selected two different GPT versions to compare whether the newer version has a significant performance increase.
For DeepSeek-R1 and Mistral-7B, we selected 7B-parameter variants to reflect more realistic deployment scenarios and to assess the performance of smaller-scale models.
Specifically, we used DeepSeek-R1-Distill-Qwen-7B and Mistral-7B-Instruct-v0.3.
For DeepSeek-R1, all responses are generated using a dedicated \emph{think mode} allowing deeper reasoning.
Each model provides a sufficiently large context window for all prompt variations.
To improve reproducibility for the local models (DeepSeek-R1, Mistral-7B), we disabled sampling and used a fixed seed (1234). 
\subsection{Prompt Design}
\label{sec:mcq_answering_prompting}
\paragraph{Prompting Strategy:}
Given the strong influence of prompt design on model performance~\cite{prompting_explanation,prompting_survey}, 
we employ a structured prompting strategy.
Several approaches and extra information can be used to improve the processing of the models~\cite{prompting_survey}.
Additionally, there are formulation guidelines to improve LLMs' processing~\cite{promptingguide,openai_prompting}.
Drawing on these guidelines, we derived the following prompt components for our structured prompting strategy:
\begin{itemize}
\item \textbf{Defined Persona}: guides the model toward a specific perspective
\item \textbf{Thematic Focus}: ensures that irrelevant content is excluded
\item \textbf{Background Knowledge}: provides additional information on the topic so the model can better process the context
\item \textbf{Level of Detail of the Response}: controls whether the model provides only answers or also explanations
\item \textbf{Task}: explicitly defines what the model is required to do 
\item \textbf{Output Format}: specifies the format for an easier evaluation
\end{itemize}
We excluded other techniques, such as few-shot learning and chain-of-thought prompting, as they were not suited to our evaluation.
Speech style, follow-up prompts, dialogue, explicit constraints, and key-term definitions were omitted to minimize stylistic and interaction-related biases.
Audience specification was implicitly integrated through persona.
\paragraph{Defined Prompts:}
Each prompt is structured into four sections: \emph{Persona}, \emph{Task Definition}, \emph{Output Format}, and \emph{Background Knowledge}.
To separate instructions from task content, the question and optional background knowledge were provided as a user prompt, while the instructions were given as a system prompt.

We evaluated three persona settings: \emph{none}, \emph{student}, and \emph{professor} to evaluate their effect on model performance.
Prior research has shown that the use of expert personas can influence outputs~\cite{persona_prompting,pattern_persona}.
We also included topic context to provide more details.
For \emph{none}, we skipped the persona section.

After the persona, the task is defined in detail to ensure correct processing.
The general topic (IVDA lecture) and the structure of the MCQs are clarified.
Two output format settings were used: \emph{answer-only} and \emph{answer-with-explanation}, both enforcing a fixed format for consistency.
To assess the impact of background knowledge, we compared prompts with and without slide-based context, manually extracted from the lecture slide set from which the current question was created.

The combination of all settings results in $12$ distinct prompts, allowing analysis of optimal configurations and influencing factors.
For the user study on the impact of LLM-generated suggestions on learner response behavior, we created a separate prompt that mirrors student prompting behavior.
Students tend to ask procedural questions without completely structuring the prompt~\cite{mcnichols2026studychatdatasetanalyzingstudent,neagu2026howiprocedural}.
To approximate student-like requests, we limited the prompt to persona, task, and brief explanations.

\subsection{User Study Design}
\label{method:user_study}
\begin{figure}[htbp]
    \centering
    \includegraphics[width=0.68\linewidth]{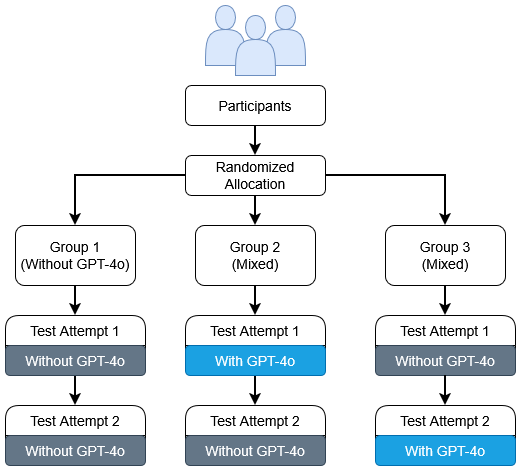}
    \caption{User study design with three randomized groups. All participants had to complete two test attempts with identical tasks under different settings: with/out GPT-4o.}
    \label{fig:study_design}
\end{figure}
We conducted a user study to analyze how pre-computed GPT-4o output influences Master's students' behavior during an MCQ test.
An overview of the user study design is given in Figure~\ref{fig:study_design}.
Participants were recruited from two specific Master-level lectures, IVDA and CV, at a German university.
Both lectures were held in English.
All students of the courses were informed about the experimental setting of the exercise and were free to choose whether to participate; approximately 30\% of the students consented to take part.
All data were fully anonymized using participant IDs to track attempts without identifying individuals.
The study was designed to reflect the participants’ level of acquired knowledge, such that the topics covered in the test had already been presented in the specific lecture.
The user study scenarios were conducted independently for each lecture, involving different participants in each case.
Participants were randomly assigned to one of three groups, while ensuring a balanced distribution across groups.
They were instructed to complete two test attempts to assess potential changes in responses.
Between the two attempts, participants were encouraged to take a break of one hour for reflection. 
The first group completed an MCQ test without GPT-4o output, the second group received the output for all questions on the first attempt, and the third group completed the test first without and then with the output.
GPT-4o output included labels for each multiple-choice option (true/false) and provided explanations, enabling students to reflect and compare with their own knowledge.
Additionally, a group-specific questionnaire assessed students’ perceptions of LLMs and the test setting.
It consisted of five sections: \emph{general use of LLMs}, \emph{personal perception}, \emph{LLM usage during the test}, \emph{acceptance of LLMs}, and \emph{feedback} to assess subjective experiences and willingness to integrate LLMs into learning.
Most questions used five-point Likert scales ranging from 1 (strongly disagree) to 5 (strongly agree) to assess students' views.
The feedback question consists of an open-text response for further insights.
Versions varied slightly by group to reflect the setting.
\subsection{Evaluation Methods}
\label{method:evaluation}
We assessed model performance across all prompt variants using multiple metrics. For scoring, each MCQ comprised four answer options worth one point each, yielding a maximum score of four points per question. 
We report two complementary measures: the mean score (total points earned out of four) and the proportion of fully correct answers - that is, the percentage of MCQs for which a model selected all correct options.
To compare model and student performance, we averaged scores per question and ranked questions by difficulty (i.e., the proportion of correct answers). 
We then examined whether models and students agreed on which questions were easiest and hardest by comparing the overlap in their top and bottom 10\,\% of questions. 
Finally, for each MCQ, we identified the most frequent incorrect option chosen by students and checked whether a model selected the same distractor, using a binary option-level similarity measure.

To statistically evaluate the user study results (see Sec.~\ref{exp:user_study}), we employed a linear mixed model (LMM), which is robust to unbalanced data and repeated measurements~\cite{lmm}. 
In our study, the experimental groups differed in the number of participants, resulting in an unbalanced dataset.
As the dependent variable, we focused on each student's score per attempt.
This forms the group, attempt, and interaction (group x attempt) as fixed effects.
Participant identity was included as a random intercept to account for repeated measurements and inter-individual differences in baseline performance.
The model structure is given in Equation~\ref{eq:lmm}.
\begin{equation}
\text{Score} \sim \text{Group} * \text{Attempt} + (1 \mid \text{Participant})
\label{eq:lmm}
\end{equation}
Based on this, the following null hypotheses were formulated:
\begin{enumerate}
    \item[\textbf{H1}] Attempt Effect: Student performance does not differ between the first and the second test attempt.
    \item[\textbf{H2}] Group Effect: Student performance does not differ between the experimental groups.
    \item[\textbf{H3}] Interaction Effect: The change in student performance between the first and second attempt does not differ between the experimental groups.
\end{enumerate}
Each of the hypotheses corresponds to the significance of a specific fixed effect in the linear mixed model.
The significance of these effects was assessed using likelihood-ratio tests (LRT), since Wald $\chi^2$ tests can be unreliable on small samples~\cite{significance_lmm}.
Still, we reported the results of such tests for comparison.
We used a conventional $\alpha = 0.05$ for evaluation.
\section{LLM-Based Chatbot Performance on MCQs}
\label{exp:answer_analysis}
\paragraph{Performance Evaluation:}
\begin{table}[t]
\centering
\caption{Results (\%) for all evaluated models across prompt variations.
Each prompt variant is identified by a three-part code: the assigned persona (\textit{None} = no persona, \textit{Stud.} = student, \textit{Prof.} = professor), whether an answer justification was requested (\textit{Y}/\textit{N}), and whether slide context was included in the prompt (\textit{Y}/\textit{N}). For example, \textit{Prof.-Y-Y} denotes a professor persona with justification requested and slide context provided. \textit{Acc.} denotes mean score (out of 4 points per question), \textit{Corr.} the percentage of fully correct answers, and \textit{Sim.} the similarity between the model's most frequent incorrect answer choice and the most frequently chosen incorrect answer by students.}
\begin{tabular}{l 
c c c | 
c c c | 
c c c | 
c c c}
\hline
 & 
\multicolumn{12}{c}{\textbf{Models}} \\
\cline{2-13}
&
\multicolumn{3}{c|}{\textbf{Mistral-7B}} &
\multicolumn{3}{c|}{\textbf{DeepSeek-R1}} &
\multicolumn{3}{c|}{\textbf{GPT-4o}} &
\multicolumn{3}{c}{\textbf{GPT-5.2}} \\
\cline{2-13}
\textbf{Setting} &
\textbf{Acc.} & \textbf{Corr.} & \textbf{Sim.} &
\textbf{Acc.} & \textbf{Corr.} & \textbf{Sim.} &
\textbf{Acc.} & \textbf{Corr.} & \textbf{Sim.} &
\textbf{Acc.} & \textbf{Corr.} & \textbf{Sim.} \\
\hline
None-N-N &
\textbf{74.6} & 28.6 & 32.0 &
79.6 & 44.3 & 33.3 &
83.9 & 57.1 & 43.3 &
87.9 & 61.4 & 55.6 \\

None-N-Y &
63.9 & 17.1 & 12.1 &
72.1 & 32.9 & 21.3 &
90.0 & 67.1 & 52.2 &
94.6 & 84.3 & \textbf{72.7} \\

None-Y-N &
73.2 & 28.6 & 40.0 &
79.6 & 40.0 & 33.3 &
84.6 & 54.3 & 50.0 &
85.4 & 54.3 & 50.0 \\

None-Y-Y &
63.9 & 24.3 & 17.0 &
66.1 & 32.9 & 10.6 &
\textbf{91.4} & 70.0 & 57.1 &
95.0 & 84.3 & 54.6 \\
\hline
Stud.-N-N &
72.9 & \textbf{31.4} & 25.0 &
75.4 & 35.7 & 33.3 &
82.5 & 50.0 & \textbf{60.0} &
87.9 & 50.0 & 46.4 \\

Stud.-N-Y &
67.5 & 25.7 & 11.5 &
73.2 & 40.0 & 14.3 &
89.6 & 68.6 & 54.6 &
95.7 & 84.3 & 63.6 \\

Stud.-Y-N &
73.2 & 28.6 & 30.0 &
80.0 & 42.9 & 37.5 &
84.3 & 55.7 & 54.8 &
88.6 & 61.4 & 51.9 \\

Stud.-Y-Y &
63.2 & 21.4 & 9.1 &
70.0 & 34.3 & 17.4 &
91.1 & \textbf{71.4} & 55.0 &
\textbf{96.4} & \textbf{87.1} & 66.7 \\
\hline
Prof.-N-N &
73.9 & 30.0 & 32.7 &
79.3 & 38.6 & \textbf{44.2} &
83.2 & 52.9 & 51.5 &
88.6 & 65.7 & 41.7 \\

Prof.-N-Y &
68.6 & 25.7 & 13.5 &
74.3 & 38.6 & 20.9 &
91.1 & \textbf{71.4} & 45.0 &
95.0 & 85.7 & 60.0 \\

Prof.-Y-N &
73.2 & 28.6 & \textbf{42.0} &
\textbf{81.8} & \textbf{45.7} & 39.5 &
82.5 & 51.4 & 50.0 &
85.7 & 55.7 & 48.4 \\

Prof.-Y-Y &
62.5 & 21.4 & 18.2 &
71.4 & 31.4 & 16.7 &
88.6 & 65.7 & 41.7 &
93.9 & 81.4 & 53.9 \\
\hline
\textbf{Avg.} &
69.2 & 26.0 & 23.6 &
75.2 & 38.1 & 26.9 &
86.9 & 61.3 & 51.3 &
91.2 & 71.3 & 55.5 \\
\hline
\end{tabular}
\label{tab:all_models_prompt_results}
\end{table}

For each prompt variation, we tested the models on all $70$ MCQs and calculated the accuracy of correctly labeled options, the percentage amount of fully correct questions, and the similarity to the average wrong student response.
Table~\ref{tab:all_models_prompt_results} reports the results for all investigated models.

For Mistral-7B, accuracy varies notably across prompts.
Providing slide content constantly reduces the accuracy to $62-69\,\%$.
Approaches without content reach over $70\,\%$, suggesting larger input hinders identification of question-relevant information.
Using personas and asking for an explanation reduces accuracy by up to $1.7\,\%$. 
The highest accuracy ($\approx75\,\%$) is achieved without personas, explanations, and slide context.
Across all variants, about $70\,\%$--$80\,\%$ of questions were not answered fully correctly.
The Model–student error similarity varies, indicating differences in how the information is handled.

DeepSeek-R1 performs similar to Mistral-7B, but 
shows a performance advantage.
The results differ across prompt variants, and no systematic performance pattern can be identified.
In contrast, the GPT models achieve higher accuracy and more fully correct responses, likely due to their larger scale and training.
Their performance also improved when slide content was provided.
In general, more than half of all questions are completely correct, indicating deeper processing, with GPT-5.2 outperforming GPT-4o.
The similarity error varies and reaches up to $60\,\%$ (GPT-4o) and $\approx73\,\%$  (GPT-5.2), while the smaller models generally score lower.
Still, the error similarity scores of the GPT models do not fully align with student error responses (\textbf{RQ3}).
The smaller models achieve moderate performance, 
but select incorrect options for most questions. In contrast, GPT models perform better (\textbf{RQ1}).
The consistently high proportion of partially correct answers and the sensitivity to context highlight current limitations of smaller LLM-based chatbots in reliably solving IVDA-MCQs.
Furthermore, the different prompting approaches 
significantly influence the results,
while no consistent effect is observed for personas (\textbf{RQ2}).
\paragraph{Chatbot-User Comparison:}
\begin{table}[t]
\centering
\caption{Top and bottom 10\,\% of questions based on the average performance of models and students. Each entry lists the question ID and percentage of correct answers in parentheses. Questions in bold appear in both the student and model rankings.}
\begin{tabular}{l|l|l}
\textbf{Rank} & \textbf{Average of Models} & \textbf{Average of Students} \\
\hline
\textbf{Top} 
& Q66 (97.9), Q30 (97.9), Q2 (95.8)
& Q69 (100.0), Q5 (97.1), Q25 (96.0) \\
& \textbf{Q50} (91.7), Q41 (89.6), \textbf{Q7} (87.5)
& Q24 (95.8), \textbf{Q50} (94.7), \textbf{Q7} (92.3) \\
& Q4 (87.5) 
& Q31 (91.7) \\
\hline
\textbf{Worst} 
& Q62 (0.0), \textbf{Q40} (6.3), \textbf{Q47} (6.3)
& Q26 (0.0), Q27 (0.0), \textbf{Q63} (0.0) \\
& \textbf{Q63} (10.4), Q45 (12.5), Q48 (12.5)
& Q28 (4.2), \textbf{Q40} (14.3), \textbf{Q47} (14.3) \\
& Q3 (16.7)
& Q61 (22.2) \\
\hline
\end{tabular}
\label{tab:question_performance}
\end{table}

Table~\ref{tab:question_performance} shows limited consistency between the best- and worst-performing questions for models and students, with minimal overlap of questions.
Considering the performance score, both exhibit comparable behavior: the best-performing questions achieve similarly accurate scores above $85\,\%$, while the worst-performing ones often fall below $15\,\%$.
This indicates comparable difficulty distributions, reflecting similar average performance for top and worst questions rather than identical error characteristics (\textbf{RQ3}).
\paragraph{Stability Evaluation:}
We conducted a stability evaluation because the models are not fully deterministic.
We focus on the smaller models, as they are more easily reproducible and configurable locally.
We tested robustness to option ordering by rotating the positions of the MCQ options. 
For each question, we generate four permutations by cyclically shifting the options, ensuring that each option appears in every position exactly once.
We further analyze the effect of randomness when sampling is enabled (default sampling settings) by running the identical prompt $10$ times.
For both models, we used the prompt with the highest accuracy to focus only on robustness.
As shown in Table~\ref{tab:stability_results}, varying the order of the options slightly affects the score, indicating a minor position bias for both models.
The error similarity also indicates that the models vary in terms of the wrong answers, hinting at partial guessing.
Repeating the same prompt without permuting the options (last row) leads to slight variation in terms of the scores.
Overall, accuracy and fully correct answers remain relatively stable.
The error similarity varies like the permutation evaluation, indicating the same reason.
\begin{table}[htbp]
\centering
\caption{Results (\%) for local models under option rotation (top) and 10 runs using sampling (bottom).
\textit{Acc.} denotes accuracy, \textit{Corr.} fully correct answers, and
\textit{Sim.} the error similarity to the most frequently chosen incorrect answer by students.}
\begin{tabular}{l c c c | c c c}
\hline
 &
\multicolumn{3}{c|}{\textbf{Mistral-7B}} &
\multicolumn{3}{c}{\textbf{DeepSeek-R1}} \\
\cline{2-7}
\textbf{Setting} & \textbf{Acc.} & \textbf{Corr.} & \textbf{Sim.} 
& \textbf{Acc.} & \textbf{Corr.} & \textbf{Sim.} \\
\hline
Permutation: 0123 & \textbf{74.6} & 28.6 & \textbf{32.0} & \textbf{81.8} & \textbf{45.7} & 39.5 \\
Permutation: 3012 & 69.6 & 25.7 & 23.0 & 77.9 & 37.1 & 36.4 \\
Permutation: 2301 & 70.7 & 25.7 & 19.2 & 75.7 & 40.0 & 42.5 \\
Permutation: 1230 & 72.9 & \textbf{32.9} & 25.5 & 78.6 & 37.1 & \textbf{43.2} \\
Std. & $\pm 2.2$ & $\pm 3.4$ & $\pm 5.4$ & $\pm 2.5$ & $\pm 4.0$ & $\pm 3.1$ \\
\hline
Std. (10 runs: 0123)
& $\pm 0.9$ & $\pm 2.4$ & $\pm 4.1$ 
& $\pm 1.6$ & $\pm 3.0$ & $\pm 4.0$ \\

\hline
\end{tabular}
\label{tab:stability_results}
\end{table}
\section{GPT Impact on Student Responses}
\label{exp:user_study}
\paragraph{Test Results:}
\begin{table}[t]
\centering
\caption{Comparison of participant groups and \textit{GPT-4o}. \textit{N} describes number of participants, \textit{Min.}, \textit{Max.} and \textit{Avg.} refers to the points of the students, \textit{Avg. \%} denotes average performance, \textit{Sim.} the similarity to \textit{GPT-4o} outputs, and \textit{Better/Same} the proportion of students outperforming or matching \textit{GPT-4o}. \textit{G}: group; \textit{A}: attempt.}
\begin{tabular}{l l r r r r r r r r r}
\hline
\textbf{Setting} & \textbf{Group} & \textbf{N} & \textbf{Min.} & \textbf{Max.} & \textbf{Avg.} & \textbf{Avg. \%} & \textbf{Sim.} & \textbf{Better} & \textbf{Same} & \textbf{Worse}\\
\hline
IVDA & G1-A1 & 15 & 14 & 19 & 15.86 & 79.29 & 64.76 & 4.67 & 9.52 & 85.81\\
IVDA & G1-A2 & 15 & 14 & 19 & 16.13 & 80.67 & 65.00 & 6.67 & 13.33 & 80.00\\
IVDA & G2-A1 & 10 & 15 & 18 & 17.00 & 85.00 & \textbf{87.50} & 0.00 & \textbf{50.00} & \textbf{50.00}\\
IVDA & G2-A2 & 10 & 14 & 18 & 16.70 & 83.50 & 79.50 & 0.00 & 40.00 & 60.00\\
IVDA & G3-A1 & 11 & 12 & 19 & 16.18 & 80.90 & 64.09 & \textbf{9.09} & 0.00 & 90.91\\
IVDA & G3-A2 & 11 & 15 & \textbf{20} & 17.09 & 85.46 & 80.91 & \textbf{9.09} & 18.18 & 72.73\\ 
IVDA & GPT-4o & – & \textbf{18} & 18 & \textbf{18.00} & \textbf{90.00} & – & – & – & -\\\hline

CV & G1-A1 & 16 & 23 & \textbf{32} & 29.25 & 91.41 & 82.03 & 87.50 & 0.00 & 12.50\\
CV & G1-A2 & 16 & 22 & \textbf{32} & 29.38 & 91.80 & 81.64 & 87.50 & 0.00 & 12.50\\
CV & G2-A1 & 9 & 21 & \textbf{32} & 27.56 & 86.11 & 85.76 & 55.56 & \textbf{11.11} & 33.33\\
CV &G2-A2 & 9 & 25 & 31 & 28.78 & 89.93 & \textbf{91.67} & 66.67 & 2.22 & 31.11\\
CV & G3-A1 & 9 & \textbf{29} & \textbf{32} & \textbf{30.33} & \textbf{94.79} & 86.81 & \textbf{100.00} & 0.00 & \textbf{0.00}\\
CV & G3-A2 & 9 & 27 & \textbf{32} & 30.00 & 93.75 & 90.63 & 89.88 & \textbf{11.11} & \textbf{0.00}\\
CV & GPT-4o & – & 27 & 27 & 27.00 & 84.38 & – & – & – & -\\
\hline
\end{tabular}
\label{tab:group_gpt_comparison}
\end{table}

Due to incomplete participation, group sizes varied, and only students completing both attempts were included to ensure comparability.
Table~\ref{tab:group_gpt_comparison} summarizes results across groups, attempts, and GPT-4o.
For IVDA-MCQs, GPT-4o generally outperforms an average student.
Still, there are students with at least an equivalent score, although this occurs infrequently in most attempts, except in Group~2.
Group~1 achieves the lowest average score across both attempts, while Group~3 aligns with Group~1 in the first attempt and Group~2 in the second.
Group~2 performed slightly worse during the second attempt, implying that they were uncertain after seeing the questions without the output from GPT-4o.
Similarities between student and GPT-4o solutions are also comparable to the test setting, indicating frequent use of solutions (\textbf{RQ4}). 

In the CV scenario, the average student outperforms GPT in all attempts, with many achieving comparable or higher scores.
The similarity to GPT-4o output is always over $80\,\%$. 
Group~1 remains stable across attempts.
Group~2 improves slightly, while Group~3 declines in the second attempt.
This suggests that GPT-4o outputs may reduce performance when prior knowledge is higher, potentially introducing uncertainty rather than clarification (\textbf{RQ4}).
\begin{table*}[t]
\centering
\caption{Change in student performance (\%) between the first and second attempt of each group in both user study settings. 
\textit{N} describes the number of participants.
\textit{Improved}, \textit{Unchanged}, and \textit{Worsened} indicate performance changes; \textit{Better than GPT} denotes participants outperforming GPT-4o in both attempts. \textit{G}: group; \textit{A}: attempt.}
\begin{tabular}{l l c c c c | c}
\hline
\textbf{Setting} & \textbf{Group} & \textbf{N} & \textbf{Improved} & \textbf{Unchanged} & \textbf{Worsened} & \textbf{Better than GPT} \\
\hline
IVDA & G1 A1→A2 & 15 & 13.33 & \textbf{80.00} & 6.67 & 6.67 \\
IVDA & G2 A1→A2 & 10 & 20.00 & 60.00 & \textbf{20.00} & 0.00 \\
IVDA & G3 A1→A2 & 11 & \textbf{63.64} & 18.18 & 18.18 & \textbf{9.09} \\
\hline
CV & G1 A1→A2 & 16 & 25.00 & 50.00 & \textbf{25.00} & 81.25 \\
CV & G2 A1→A2 & 9  & \textbf{33.33} & 55.56 & 11.11 & 55.56 \\
CV & G3 A1→A2 & 9  & 0.00  & \textbf{77.78} & 22.22 & \textbf{88.89} \\
\hline
\end{tabular}
\label{tab:short_attempt_comparison}
\end{table*}

Table~\ref{tab:short_attempt_comparison} summarizes changes between the two attempts. 
In the IVDA scenario, $63.64\,\%$ of Group~3 improved, while $18.18\,\%$ performed worse. 
In particular, $\approx80\,\%$ of students in Group~1 and Group~2 show unchanged performance, indicating that the altered test setting in Group~3 has an observable effect.
Only a few students outperform GPT in both attempts, suggesting reliance on GPT and potential uncertainty (\textbf{RQ4}).

In the CV scenario, at least half of the students achieved identical scores across the groups.
In Groups~1 and~3, over $80\,\%$ outperformed GPT-4o in both attempts.
Group~2 shows the largest gain, possibly following an initial GPT-induced deficit, while also having the fewest students outperforming GPT across both attempts.
This suggests a possible trend that, depending on prior knowledge, GPT-generated outputs may influence students' behavior differently (\textbf{RQ4}).
\paragraph{Statistical Evaluation:}
\begin{table}[t]
\centering
\caption{p-values of LRT and Wald tests for hypotheses (H) in both user study scenarios (IVDA and CV).
Bold indicates significance at $\alpha = 0.05$.}
\begin{tabular}{lcc|cc}
\hline
\textbf{H} & \textbf{LRT (IVDA)} & \textbf{LRT (CV)} & \textbf{Wald (IVDA)} & \textbf{Wald (CV)}\\
\hline
\textbf{H1} & 0.229 & 0.438 & 0.650 & 0.814\\
\textbf{H2} & 0.449 & 0.145 & 0.277 & \textbf{0.041}\\
\textbf{H3} & 0.058 & 0.285 & \textbf{0.046} & 0.272\\
\hline
\label{tab:statistic_results}
\end{tabular}
\end{table}
Table~\ref{tab:statistic_results} shows the p-values of the statistical tests for both scenarios.
Based on LRT, no fixed effects are significant at $\alpha = 0.05$, suggesting no attempt-dependent performance differences between experimental conditions.
For IVDA-MCQs, the interaction effect (\textbf{H3}) approaches significance ($p=0.058$), indicating a tendency toward different performance changes between the groups, as reflected in Group~3.
Under the Wald $\chi^2$ tests, the interaction effect (\textbf{H3}) for IVDA-MCQs and the group effect (\textbf{H2}) for CV-MCQs are significant.
Overall, no conclusive evidence is found for a significant impact of GPT-generated output.
However, the tendency for \textbf{H3} in IVDA-MCQs and the limited sample size 
motivate larger studies to gain reliable insights (\textbf{RQ4}).
\paragraph{Questionnaire Results:}
Questionnaire responses provide insights into the students' thoughts (\textbf{RQ4}).
Across both user study scenarios, students report regular use of LLMs for studying. 
When GPT output was given, it was rated as understandable and helpful, often increasing confidence and engagement.
However, trust varied across the scenarios.
For IVDA-MCQs, agreement was neutral to moderate, whereas for CV-MCQs, trust was more skeptical.
This suggests that students see utility but do not unconditionally trust these systems.
The open-answer feedback on how helpful and problematic such models are shows that they were perceived as helpful for simplifying concepts and improving learning efficiency, but problematic due to hallucinations, overconfidence, and inconsistencies in both user study scenarios. 
The IVDA-MCQs scenario highlighted risks of increased dependency and cognitive offloading, while the other scenario emphasized organizational benefits alongside answer uncertainty.
Overall, students view these models as useful for learning but requiring critical use (\textbf{RQ4}).
\paragraph{Pedagogical and Socio-Technical Aspects:}
Another important aspect is the discussion of the overall results from a pedagogical and socio-technical perspective.
Pedagogically, our results indicate a first trend that LLM-generated output can serve as scaffolding, particularly supporting students with lower prior knowledge, who benefit more from structured guidance due to higher cognitive load~\cite{ngu2023instructional}.
In this context, these models can be seen as a technological layer linking pedagogical guidance and domain knowledge, which aligns with the TPACK framework~\cite{tpack}.
However, results for CV-MCQs indicate that for students with higher prior knowledge, LLM output may negatively affect performance, suggesting limited benefits and increased cognitive offloading, i. e., relying on the output instead of reasoning.
From the socio-technical perspective, LLM-based chatbots allow interaction.
Learners can, e.g., discuss their question in more detail if the output was unclear or they want more information about it, like in a social interaction.
However, the questionnaire results indicate that students have limited trust compared to human support.
Still, this aligns with the community of inquiry framework (CoI)~\cite{Micsky20102019}, as aspects of teaching presence can be simulated.
\section{Conclusions}
\label{sec:conclusions}
In this paper, we have analyzed the performance of LLM-based outputs on university-level MCQs and examined the influence on student behavior in a user study.
The results show that smaller models struggle to reliably answer MCQs in a zero-shot setting.
In contrast, GPT-based models achieve higher accuracy, although errors remain.
This suggests that models require fine-tuning to course-specific material for educational use (\textbf{RQ1}).
Prompt design affects performance, with smaller models showing reduced accuracy when additional context is provided, while GPT-based models benefit from it.
Still, persona effects remain inconsistent (\textbf{RQ2}).
Model and student responses exhibit limited alignment at the question level, yet show comparable performance distributions across easy and difficult questions (\textbf{RQ3}).
LLM-based models could be trained on student error patterns to support difficulty-based question classification.
Looking at the user study, the quantitative results indicate an influence of chatbot-generated outputs on student performance, but this effect depends on both the task domain and the experimental setup. 
Depending on the task, chatbot outputs can support or hinder learning, highlighting the need for careful educational integration (\textbf{RQ4}).
However, these findings are limited by our dataset and the small group of participants. 

Future work could extend the analysis to larger datasets 
across CS domains and difficulty levels (e.g., secondary to graduate level). 
In addition, a larger user study with more participants and questions would enable a more reliable assessment of the statistical significance of chatbot-generated output effects.

\begin{credits}
\subsubsection{\ackname} This work has been financially supported by the Lower Saxony Ministry of Science and Culture with funds from the zukunft.niedersachsen program of the VolkswagenStiftung for the project "VidQA: Automated comprehension tests for learning videos".

\subsubsection{\discintname}
The authors have no competing interests to declare that are relevant to the content of this article.
\end{credits}
%
%
%
\bibliographystyle{splncs04}
\bibliography{references}

@inproceedings{persona_prompting,
  title={Evaluating persona prompting for question answering tasks},
  author={Olea, Carlos and Tucker, Holly and Phelan, Jessica and Pattison, Cameron and Zhang, Shen and Lieb, Maxwell and Schmidt, Doug and White, Jules},
  booktitle={International conference on artificial intelligence and soft computing, AIS (2024), Sydney, Australia, June 22-23, 2024},
  year={2024},
  doi={10.5121/csit.2024.141106}
}

@inproceedings{chatgpt_chances_education,
  author       = {Hyanghee Park and
                  Daehwan Ahn},
  editor_       = {Florian 'Floyd' Mueller and
                  Penny Kyburz and
                  Julie R. Williamson and
                  Corina Sas and
                  Max L. Wilson and
                  Phoebe O. Toups Dugas and
                  Irina Shklovski},
  title        = {The Promise and Peril of ChatGPT in Higher Education: Opportunities,
                  Challenges, and Design Implications},
  booktitle    = {Conference on Human Factors in Computing
                  Systems, {CHI} 2024, Honolulu, HI, USA, May 11-16, 2024},
  pages        = {271:1--271:21},
  publisher    = {{ACM}},
  year         = {2024},
  url_          = {https://doi.org/10.1145/3613904.3642785},
  doi          = {10.1145/3613904.3642785},
  bibsource    = {dblp computer science bibliography, https://dblp.org}
}

@article{llm_chances_education,
  author       = {Shen Wang and
                  Tianlong Xu and
                  Hang Li and
                  Chaoli Zhang and
                  Joleen Liang and
                  Jiliang Tang and
                  Philip S. Yu and
                  Qingsong Wen},
  title        = {Large Language Models for Education: {A} survey and outlook},
  journal      = {{IEEE} Signal Processing Magazine},
  volume       = {42},
  number       = {6},
  pages        = {51--63},
  year         = {2025},
  url_          = {https://doi.org/10.1109/MSP.2025.3594309},
  doi          = {10.1109/MSP.2025.3594309},
  bibsource    = {dblp computer science bibliography, https://dblp.org}
}

@inproceedings{potential_learning_outcomes,

  author={Alnajjar, Fady and Alneyadi, Alreem R. and Almetnawy, Habiba and Alneyadi, Khawla S. and Alhemeiri, Fatima and Alneyadi, Amna R. and Orabi, Ahed and Palliyalil, Muhammed S.},
  title={Exploring the Pedagogical Potential of Large Language Models: A Multimodal Study on Student Learning Outcomes}, 
  booktitle={International Workshop on Artificial Intelligence and Education, {WAIE} 2024, Tokyo, Japan, August 28-30, 2024}, 
  year={2024},
  publisher={IEEE},
  volume_={},
  number_={},
  pages={93-96},
  doi={10.1109/WAIE63876.2024.00024}
}

@article{prompting_survey,
  author       = {Pranab Sahoo and
                  Ayush Kumar Singh and
                  Sriparna Saha and
                  Vinija Jain and
                  Samrat Mondal and
                  Aman Chadha},
  title        = {A Systematic Survey of Prompt Engineering in Large Language Models:
                  Techniques and Applications},
  journal      = {arXiv preprint},
  volume       = {abs/2402.07927},
  year         = {2024},
  url_          = {https://doi.org/10.48550/arXiv.2402.07927},
  doi          = {10.48550/ARXIV.2402.07927},
  eprinttype    = {arXiv},
  eprint       = {2402.07927},
  bibsource    = {dblp computer science bibliography, https://dblp.org}
}

@inproceedings{pattern_persona,
author = {Schreiber, William and White, Jules and Schmidt, Douglas C.},
title = {Toward A Pattern Language for Persona-Based Interactions with LLMs},
year = {2026},
isbn_ = {9781941652206},
publisher = {The Hillside Group},
address_ = {USA},
url_ = {https://doi.org/10.64346/PLoP2024p27},
doi = {10.64346/PLoP2024p27},
booktitle = {Conference on Pattern Languages of Programs, People, and Practices, {PLoP} 2024, Columbia River Gorge, WA, USA, October 13 - 16, 2024},
articleno_ = {23},
numpages = {19},
location_ = {Columbia River Gorge, WA, USA},
series_ = {PLoP '24}
}

@article{school_level_questions,
  author       = {Subhankar Maity and
                  Aniket Deroy and
                  Sudeshna Sarkar},
  title        = {Can large language models meet the challenge of generating school-level questions?},
  journal      = {Computers and Education: Artificial Intelligence},
  volume       = {8},
  pages        = {100370},
  year         = {2025},
  url_          = {https://doi.org/10.1016/j.caeai.2025.100370},
  doi          = {10.1016/J.CAEAI.2025.100370},
  bibsource    = {dblp computer science bibliography, https://dblp.org}
}

@inproceedings{bloom_qg,
  author       = {Nicy Scaria and
                  Suma Dharani Chenna and
                  Deepak N. Subramani},
  editor       = {Andrew M. Olney and
                  Irene{-}Angelica Chounta and
                  Zitao Liu and
                  Olga C. Santos and
                  Ig Ibert Bittencourt},
  title        = {Automated Educational Question Generation at Different Bloom's Skill Levels Using Large Language Models: Strategies and Evaluation},
  booktitle    = {Artificial Intelligence in Education,
                  {AIED} 2024, Recife, Brazil, July 8-12, 2024},
  series_       = {Lecture Notes in Computer Science},
  volume_       = {14830},
  pages        = {165--179},
  publisher    = {Springer},
  year         = {2024},
  url_          = {https://doi.org/10.1007/978-3-031-64299-9\_12},
  doi          = {10.1007/978-3-031-64299-9\_12},
  bibsource    = {dblp computer science bibliography, https://dblp.org}
}

@article{llm_tutor_review,
  author       = {Silvia Garc{\'{\i}}a{-}M{\'{e}}ndez and
                  Francisco de Arriba{-}P{\'{e}}rez and
                  Maria del Carmen Lopez{-}Perez},
  title        = {A review on the use of large language models as virtual tutors},
  journal      = {Science \& Education},
  volume       = {34},
  pages         = {877--892},
  year          = {2025},
  url_          = {https://doi.org/10.1007/s11191-024-00530-2},
  doi          = {10.1007/s11191-024-00530-2},
}

@article{review_llm_education,
  title={Pedagogical alignment of large language models (llm) for personalized learning: a survey, trends and challenges},
  author={Razafinirina, Mahefa Abel and Dimbisoa, William Germain and Mahatody, Thomas},
  journal={Journal of Intelligent Learning Systems and Applications},
  volume={16},
  number={4},
  pages={448--480},
  year={2024},
  doi={10.4236/jilsa.2024.164023},
  publisher={Scientific Research Publishing}
}

@article{quiz_generation_gpt,
title = {Exploring the potential of ChatGPT to create multiple-choice question exams},
journal = {Educación Médica},
volume = {25},
number = {4},
pages = {100930},
year = {2024},
issn = {1575-1813},
doi = {10.1016/j.edumed.2024.100930},
url_ = {https://www.sciencedirect.com/science/article/pii/S1575181324000457},
author = {Cristian N. Rivera-Rosas and J.R. Tadeo Calleja-López and Enrique Ruibal-Tavares and Arturo Villanueva-Neri and Cinthya M. Flores-Felix and Sergio Trujillo-López},
}

@article{gpt_problems_exam,
title = {ChatGPT 3.5 fails to write appropriate multiple choice practice exam questions},
journal = {Academic Pathology},
volume = {11},
number = {1},
pages = {100099},
year = {2024},
issn = {2374-2895},
doi = {10.1016/j.acpath.2023.100099},
url_ = {https://www.sciencedirect.com/science/article/pii/S2374289523000313},
author = {Alexander Ngo and Saumya Gupta and Oliver Perrine and Rithik Reddy and Sherry Ershadi and Daniel Remick}
}

@article{shahzad2025comprehensive,
  title={A comprehensive review of large language models: issues and solutions in learning environments},
  author={Shahzad, Tariq and Mazhar, Tehseen and Tariq, Muhammad Usman and Ahmad, Wasim and Ouahada, Khmaies and Hamam, Habib},
  journal={Discover Sustainability},
  volume={6},
  number={1},
  pages={27},
  year={2025},
  publisher={Springer},
  doi={10.1007/s43621-025-00815-8}
}

@article{jovst2024impact,
  title={The impact of large language models on programming education and student learning outcomes},
  author={Jo{\v{s}}t, Gregor and Taneski, Viktor and Karakati{\v{c}}, Sa{\v{s}}o},
  journal={Applied Sciences},
  volume={14},
  number={10},
  pages={4115},
  year={2024},
  publisher={MDPI},
  doi = {10.3390/app14104115}
}

@article{student_interaction_llm,
  author       = {Pratyusha Maiti and
                  Ashok K. Goel},
  title        = {How Do Students Interact with an LLM-powered Virtual Teaching Assistant in Different Educational Settings?},
  journal      = {arXiv preprint},
  volume       = {abs/2407.17429},
  year         = {2024},
  url_          = {https://doi.org/10.48550/arXiv.2407.17429},
  doi          = {10.48550/ARXIV.2407.17429},
  eprinttype    = {arXiv},
  eprint       = {2407.17429},
  bibsource    = {dblp computer science bibliography, https://dblp.org}
}

@article{deepseekr1,
  author       = {DeepSeek{-}AI},
  title        = {DeepSeek-R1: Incentivizing Reasoning Capability in LLMs via Reinforcement Learning},
  journal      = {arXiv preprint},
  volume       = {abs/2501.12948},
  year         = {2025},
  url_          = {https://doi.org/10.48550/arXiv.2501.12948},
  doi          = {10.48550/ARXIV.2501.12948},
  eprinttype    = {arXiv},
  eprint       = {2501.12948},
  bibsource    = {dblp computer science bibliography, https://dblp.org}
}

@article{gpt_microbiology,
title={Assessing the capability of ChatGPT in answering first-and second-order knowledge questions on microbiology as per competency-based medical education curriculum},
  author={Das, Dipmala and Kumar, Nikhil and Longjam, Langamba Angom and Sinha, Ranwir and Roy, Asitava Deb and Mondal, Himel and Gupta, Pratima},
  journal={Cureus},
  volume={15},
  number={3},
  year={2023},
  publisher={Cureus},
  doi={10.7759/cureus.36034}
}

@article{gpt_math,
  author       = {Alberto Gandolfi},
  title        = {{GPT-4} in Education: Evaluating Aptness, Reliability, and Loss of Coherence in Solving Calculus Problems and Grading Submissions},
  journal      = {International Journal of Artificial Intelligence in Education},
  volume       = {35},
  number       = {1},
  pages        = {367--397},
  year         = {2025},
  url_          = {https://doi.org/10.1007/s40593-024-00403-3},
  doi          = {10.1007/S40593-024-00403-3},
  bibsource    = {dblp computer science bibliography, https://dblp.org}
}

@article{mistral_7b,
  author       = {Albert Q. Jiang and
                  Alexandre Sablayrolles and
                  Arthur Mensch and
                  Chris Bamford and
                  Devendra Singh Chaplot and
                  Diego de Las Casas and
                  Florian Bressand and
                  Gianna Lengyel and
                  Guillaume Lample and
                  Lucile Saulnier and
                  L{\'{e}}lio Renard Lavaud and
                  Marie{-}Anne Lachaux and
                  Pierre Stock and
                  Teven Le Scao and
                  Thibaut Lavril and
                  Thomas Wang and
                  Timoth{\'{e}}e Lacroix and
                  William El Sayed},
  title        = {Mistral 7B},
  journal      = {arXiv preprint},
  volume       = {abs/2310.06825},
  year         = {2023},
  url_          = {https://doi.org/10.48550/arXiv.2310.06825},
  doi          = {10.48550/ARXIV.2310.06825},
  eprinttype    = {arXiv},
  eprint       = {2310.06825},
  bibsource    = {dblp computer science bibliography, https://dblp.org}
}

@inproceedings{llm_mistakes,
  author       = {Naiming Liu and
                  Shashank Sonkar and
                  Richard G. Baraniuk},
  editor_       = {Alexandra I. Cristea and
                  Erin Walker and
                  Yu Lu and
                  Olga C. Santos and
                  Seiji Isotani},
  title        = {Do LLMs Make Mistakes Like Students? Exploring Natural Alignments
                  Between Language Models and Human Error Patterns},
  booktitle    = {Artificial Intelligence in Education,
                  {AIED} 2025, Palermo, Italy, July 22-26, 2025},
  series_       = {Lecture Notes in Computer Science},
  volume_       = {15880},
  pages        = {364--377},
  publisher    = {Springer},
  year         = {2025},
  address_      = {Palermo, Italy},
  url_          = {https://doi.org/10.1007/978-3-031-98459-4\_26},
  doi          = {10.1007/978-3-031-98459-4\_26},
  bibsource    = {dblp computer science bibliography, https://dblp.org}
}

@inproceedings{prompting_explanation,
  author       = {Laria Reynolds and
                  Kyle McDonell},
  editor_       = {Yoshifumi Kitamura and
                  Aaron Quigley and
                  Katherine Isbister and
                  Takeo Igarashi},
  title        = {Prompt Programming for Large Language Models: Beyond the Few-Shot
                  Paradigm},
  booktitle    = {Conference on Human Factors in Computing Systems, {CHI} 2021, Yokohama, Japan, May 8-13, 2021},
  pages        = {314:1--314:7},
  publisher    = {{ACM}},
  year         = {2021},
  url_          = {https://doi.org/10.1145/3411763.3451760},
  doi          = {10.1145/3411763.3451760},
  bibsource    = {dblp computer science bibliography, https://dblp.org}
}

@article{openai2023gpt,
  author       = {OpenAI},
  title        = {{GPT-4} Technical Report},
  journal      = {arXiv preprint},
  volume       = {abs/2303.08774},
  year         = {2023},
  url_          = {https://doi.org/10.48550/arXiv.2303.08774},
  doi          = {10.48550/ARXIV.2303.08774},
  eprinttype    = {arXiv},
  eprint       = {2303.08774},
  bibsource    = {dblp computer science bibliography, https://dblp.org}
}

@article{llm_feature_survey,
  author       = {Shervin Minaee and
                  Tom{\'{a}}s Mikolov and
                  Narjes Nikzad and
                  Meysam Chenaghlu and
                  Richard Socher and
                  Xavier Amatriain and
                  Jianfeng Gao},
  title        = {Large Language Models: {A} Survey},
  journal      = {arXiv preprint},
  volume       = {abs/2402.06196},
  year         = {2024},
  url_          = {https://doi.org/10.48550/arXiv.2402.06196},
  doi          = {10.48550/ARXIV.2402.06196},
  eprinttype    = {arXiv},
  eprint       = {2402.06196},
  bibsource    = {dblp computer science bibliography, https://dblp.org}
}

@Inbook{lmm,
author="Lovric, Miodrag",
editor_="Lovric, Miodrag",
title="Linear Mixed Models: An Overview",
bookTitle="International Encyclopedia of Statistical Science",
year="2025",
publisher="Springer Berlin Heidelberg",
address="Berlin, Heidelberg",
pages="1356--1374",
isbn="978-3-662-69359-9",
doi="10.1007/978-3-662-69359-9_328",
url_="https://doi.org/10.1007/978-3-662-69359-9_328"
}

@article{significance_lmm,
  title={Evaluating significance in linear mixed-effects models in R},
  author={Steven G. Luke},
  journal={Behavior Research Methods},
  year={2016},
  volume={49},
  pages={1494 - 1502},
  url={https://api.semanticscholar.org/CorpusID:4367544}
}

@misc{promptingguide,
  title        = {Prompt Engineering Guide},
  author       = {{DAIR.AI}},
  year         = {2025},
  url          = {https://www.promptingguide.ai/},
  note         = {Accessed: Jan. 17, 2026}
}

@misc{openai_prompting,
  title   = {OpenAI Prompt Engineering Guide},
  author  = {{OpenAI}},
  year    = {2025},
  url     = {https://platform.openai.com/docs/guides/prompt-engineering},
  note    = {Accessed: Jan. 17, 2026}
}

@misc{ivda_slides,
  title  = {Slides for Interactive Visual Data Analysis},
  author = {Tominski, Christian and Schumann, Heidrun},
  year   = {2020},
  url    = {https://www.ivda-book.de/slides},
  note   = {Accessed: Jan. 17, 2026}
}

@incollection{foundations_cv_ch9,
  title     = {Introduction to Learning},
  booktitle = {Foundations of Computer Vision},
  author    = {Torralba, Antonio and Isola, Phillip and Freeman, William T.},
  chapter   = {9},
  year      = {2024},
  publisher = {MIT Press},
  url       = {https://visionbook.mit.edu/intro_to_learning.html},
  note      = {Open-access edition. Accessed: Jan. 17, 2026},
  isbn      = {978-0-262-04897-2}
}

@article{kalai2025languagemodelshallucinate,
  author       = {Adam Tauman Kalai and
                  Ofir Nachum and
                  Santosh S. Vempala and
                  Edwin Zhang},
  title        = {Why Language Models Hallucinate},
  journal      = {arXiv preprint},
  volume       = {abs/2509.04664},
  year         = {2025},
  url_          = {https://doi.org/10.48550/arXiv.2509.04664},
  doi          = {10.48550/ARXIV.2509.04664},
  eprinttype   = {arXiv},
  eprint       = {2509.04664},
  bibsource    = {dblp computer science bibliography, https://dblp.org}
}

@inproceedings{bang-etal-2025-hallulens,
  author       = {Yejin Bang and
                  Ziwei Ji and
                  Alan Schelten and
                  Anthony Hartshorn and
                  Tara Fowler and
                  Cheng Zhang and
                  Nicola Cancedda and
                  Pascale Fung},
  editor_       = {Wanxiang Che and
                  Joyce Nabende and
                  Ekaterina Shutova and
                  Mohammad Taher Pilehvar},
  title        = {HalluLens: {LLM} Hallucination Benchmark},
  booktitle    = {Association for Computational
                  Linguistics, {ACL} 2025, Vienna, Austria,
                  July 27 - August 1, 2025},
  pages        = {24128--24156},
  publisher    = {Association for Computational Linguistics},
  year         = {2025},
  url_          = {https://aclanthology.org/2025.acl-long.1176/},
  doi          = {10.18653/v1/2025.acl-long.1176},
  bibsource    = {dblp computer science bibliography, https://dblp.org}
}

@article{promptformattingimpact,
  author       = {Jia He and
                  Mukund Rungta and
                  David Koleczek and
                  Arshdeep Sekhon and
                  Franklin X. Wang and
                  Sadid Hasan},
  title        = {Does Prompt Formatting Have Any Impact on {LLM} Performance?},
  journal      = {arXiv preprint},
  volume       = {abs/2411.10541},
  year         = {2024},
  url_          = {https://doi.org/10.48550/arXiv.2411.10541},
  doi          = {10.48550/ARXIV.2411.10541},
  eprinttype   = {arXiv},
  eprint       = {2411.10541},
  bibsource    = {dblp computer science bibliography, https://dblp.org}
}

@inproceedings{robustnessreliability,
  author       = {Riccardo Lunardi and
                  Vincenzo Della Mea and
                  Stefano Mizzaro and
                  Kevin Roitero},
  editor_       = {In{\^{e}}s Lynce and
                  Nello Murano and
                  Mauro Vallati and
                  Serena Villata and
                  Federico Chesani and
                  Michela Milano and
                  Andrea Omicini and
                  Mehdi Dastani},
  title        = {On Robustness and Reliability of Benchmark-Based Evaluation of LLMs},
  booktitle    = {European Conference on Artificial Intelligence, {ECAI} 2025,Bologna, Italy, October 25-30, 2025},
  series       = {Frontiers in Artificial Intelligence and Applications},
  pages        = {4603--4610},
  publisher    = {{IOS} Press},
  year         = {2025},
  url_          = {https://doi.org/10.3233/FAIA251363},
  doi          = {10.3233/FAIA251363},
  bibsource    = {dblp computer science bibliography, https://dblp.org}
}

@article{neagu2026howiprocedural,
  author       = {Alexandra Neagu and
                  Marcus Messer and
                  Peter Johnson and
                  Rhodri Nelson},
  title        = {"How Do {I} ...?": Procedural Questions Predominate Student-LLM
                  Chatbot Conversations},
  journal      = {arXiv preprint},
  volume       = {abs/2602.18372},
  year         = {2026},
  url_          = {https://doi.org/10.48550/arXiv.2602.18372},
  doi          = {10.48550/ARXIV.2602.18372},
  eprinttype   = {arXiv},
  eprint       = {2602.18372},
  bibsource    = {dblp computer science bibliography, https://dblp.org}
}

@inproceedings{mcnichols2026studychatdatasetanalyzingstudent,
  author       = {Hunter McNichols and
                  Fareya Ikram and
                  Andrew Lan},
  title        = {The StudyChat Dataset: Analyzing Student Dialogues With ChatGPT in
                  an Artificial Intelligence Course},
  booktitle    = {International Learning Analytics and Knowledge Conference, {LAK} 2026, Bergen, Norway, April 27 - May 1, 2026},
  pages        = {53--63},
  publisher    = {{ACM}},
  year         = {2026},
  url_          = {https://doi.org/10.1145/3785022.3785029},
  doi          = {10.1145/3785022.3785029},
  bibsource    = {dblp computer science bibliography, https://dblp.org}
}

@article{Micsky20102019,
author = {Tami Micsky and Leonora Foels},
title = {Community of Inquiry (CoI): A Framework for Social Work Distance Educators},
journal = {Journal of Teaching in Social Work},
volume = {39},
number = {4-5},
pages = {293--307},
year = {2019},
publisher = {Routledge},
doi = {10.1080/08841233.2019.1642976},
URL_ = {    
        https://doi.org/10.1080/08841233.2019.1642976
},
eprint = {  
        https://doi.org/10.1080/08841233.2019.1642976  
}
}

@article{tpack,
author = {Punya Mishra and Matthew J. Koehler},
title ={Technological Pedagogical Content Knowledge: A Framework for Teacher Knowledge},

journal = {Teachers College Record},
volume = {108},
number = {6},
pages = {1017-1054},
year = {2006},
doi = {10.1111/j.1467-9620.2006.00684.x},

URL_ = { 
    
        https://doi.org/10.1111/j.1467-9620.2006.00684.x
    
    

},
eprint = { 
    
        https://doi.org/10.1111/j.1467-9620.2006.00684.x
    
    

}
}

@article{ngu2023instructional,
  title={Instructional efficiency: The role of prior knowledge and cognitive load},
  author={Ngu, Bing Hiong and Phan, Huy P and Usop, Hasbee and Hong, Kian Sam},
  journal={Applied Cognitive Psychology},
  volume={37},
  number={6},
  pages={1223--1237},
  year={2023},
  publisher={Wiley Online Library},
  doi={10.1002/acp.4117}
}
\end{document}